\documentclass[12pt]{amsart}

\newcommand{\R}{\mathbb R}

\newcommand{\eqdef}{\stackrel{\text{def}}{=}}
\newcommand{\Symb}{Symb}
\newcommand{\Id}{Id}
\newcommand{\reg}{\text{reg}}
\newcommand{\x}{\mathbf x}
\newcommand{\y}{\mathbf y}

\begin{document}

\title{On the mathematical sense of renormalization}
\author{A. V. Stoyanovsky}
\thanks{Partially supported by the grant RFBR 10-01-00536.}
\email{alexander.stoyanovsky@gmail.com}
\address{Russian State University of Humanities}

\begin{abstract}
We place the renormalization procedure in quantum field theory into the familiar mathematical context of quantization
of Poisson algebras. The Poisson algebra in question is the algebra of classical field theory Hamiltonians constructed in
a previous paper (arXiv:1008.3333). Its quantum deformations presumably contain (non-canonically)
the algebra of functional differential
operators. We explain that this picture contains renormalization as a natural ingredient.
\end{abstract}

\maketitle

\section*{Introduction}

The aim of this paper is to give a mathematical background to the notion of renormalization in quantum field theory (QFT).
Some interesting attempts to understand renormalization mathematically have been made many times, see, for example,
[1--3] and references therein. However, none of them was considered as satisfactory.
For example, the construction of Connes and Kreimer [1]
depends on a particular choice of regularization scheme, etc. We are going to give a transparent mathematical explanation
of renormalization.

The key point in our approach is the construction of the Poisson algebra of classical Hamiltonians in field theory [4].
We recall this construction and explain its naturalness in \S1. In [4] also the problem has been posed to quantize this
Poisson algebra. When restricted to the Poisson subalgebra of regular (non-local) Hamiltonians, this problem has a trivial
solution: a quantization is the algebra of functional differential operators (or the infinite dimensional Moyal algebra,
which is the same). However, when we consider also singular (local) Hamiltonians, the situation drastically changes.
The problem of quantization (or deformation quantization) becomes difficult and reminds the Kontsevich approach to
deformation quantization of Poisson manifolds [5]. This approach is closely related to QFT [6], so one can expect that
solution of our quantization problem can be achieved by QFT methods. Anyway, provided that a quantization has been
constructed, we can obtain a non-canonical embedding of the algebra of functional differential operators into the
constructed quantum algebra. However, this embedding cannot agree with taking the (non-canonical) classical symbol of a
quantum
Hamiltonian. Therefore we obtain a (non-unique) map from the space of classical non-local Hamiltonians to itself.
This map is the required renormalization map. Details are exposed in \S2.

\section{Construction of the Poisson algebra of classical Hamiltonians in field theory}

The generally covariant Hamiltonian classical field theory is exposed in detail in [7] and references therein.

Let us pose the following question. Let $\varphi=\varphi(\x)$ be the
classical (scalar for simplicity) field on a space-like surface $C$ in space-time $\R^{n+1}$,
$\x=(x_1,\ldots,x_n)$ be coordinates on the surface, and let $\pi=\pi(\x)$ be the conjugate momentum.
Both $\varphi$ and $\pi$ belong to the Schwartz space $S$ of smooth functions rapidly decreasing at infinity.
The question is: what functionals $H(\varphi,\pi)$ can be classical field theory Hamiltonians? This class of functionals
should contain the known examples, and should agree with the picture for free field (quadratic Hamiltonians).

Recall that the usual known examples of Hamiltonians are integrals over $C$ of polynomial densities depending locally
on $\varphi$ and $\pi$. On the other hand, in the case of free Klein--Gordon field, the evolution operators from
one space-like surface to another belong to the group of continuous symplectic transformations of the
symplectic topological vector space $S\oplus S$. Therefore it is natural to expect that the quadratic Hamiltonians
form the Lie algebra of this group.

It is easy to see that the answer to our question should be the following.
\medskip

{\bf Definition.}[4] A continuous polynomial functional
$H(\varphi,\pi)$ on $S\times S$ is called a Hamiltonian (or a symbol) if its first
functional differential $\delta H$, which is a linear functional
on test functions $(\delta\varphi,\delta\pi)\in S\oplus S$ for every $(\varphi,\pi)$, belongs to $S\oplus S\subset
S'\oplus S'$ (here $S'$ is the space of tempered distributions dual to $S$). Moreover, $\delta H$ should be
infinitely differentiable as an $S\oplus S$-valued functional on $S\times S$.

Denote the space of symbols by $\Symb$.
\medskip

{\bf Proposition.} $\Symb$ is a topological Poisson algebra with respect to the
standard Poisson bracket
\begin{equation}
\{H_1,H_2\}=\int\left(\frac{\delta H_1}{\delta\pi(\x)}\frac{\delta H_2}{\delta\varphi(\x)}
-\frac{\delta H_1}{\delta\varphi(\x)}\frac{\delta H_2}{\delta\pi(\x)}\right)d\x.
\end{equation}
\medskip

{\it Proof} is straightforward.
\medskip

Any $H\in\Symb$ has the form
\begin{equation}
\begin{aligned}{}
H(\varphi,\pi)&=\sum_{k=0}^N\sum_{l=0}^M H_{k,l}(\varphi,\pi),\\
H_{k,l}(\varphi,\pi)&=\frac1{k!l!}\int a_{k,l}(\x_1,\ldots,\x_k;\y_1,\ldots,\y_l)\\
&\times\varphi(\x_1)\ldots\varphi(\x_k)\pi(\y_1)\ldots\pi(\y_l)d\x_1\ldots d\x_k d\y_1\ldots d\y_l
\end{aligned}
\end{equation}
for certain tempered distributions $a_{k,l}$ symmetric in $\x_1,\ldots,\x_k$ and in $\y_1,\ldots,\y_l$ (by the Schwartz
kernel theorem).
\medskip

{\bf Definition.} If all $a_{k,l}$ are smooth functions rapidly decreasing at infinity then the Hamiltonian
$H$ is called {\it regular}. Otherwise it is called {\it singular}.
\medskip

Denote the subspace of $\Symb$ consisting of regular Hamiltonians by $\Symb^{\reg}$. Clearly,
it is a Poisson subalgebra dense in $\Symb$.

\section{The quantization problem and renormalization}

\subsection{} {\bf Definition.} A {\it quantization} ({\it deformation quantization})
of the Poisson algebra $\Symb$ is a continuous associative product
\begin{equation}
(H_1,H_2)\to H_1*H_2
\end{equation}
on the topological vector space $\Symb$ smoothly (resp. formally) depending on a parameter $h$ such that
\begin{equation}
\begin{aligned}{}
\text{ i) }& H_1*H_2=H_1H_2+O(h),\\
\text{ii) }& [H_1,H_2]\eqdef H_1*H_2-H_2*H_1=ih\{H_1,H_2\}+O(h^2),\\
\text{ iii) }& \text{if }H_1,H_2\text{ are quadratic }(k+l\le 2\text{ in (2)})\text{ then }\\
&[H_1,H_2]=ih\{H_1,H_2\}.
\end{aligned}
\end{equation}
Two (deformation) quantizations $*_1$ and $*_2$ are called {\it equivalent} if there exists a linear map
from $\Symb$ to itself smoothly (resp. formally) depending on $h$, of the form $\Id+O(h)$, which takes
$*_1$ to $*_2$.
\medskip

{\bf Problems.} Find a (deformation) quantization of $\Symb$. Classify all (deformation) quantizations
up to equivalence.

\subsection{Discussion of the quantization problem and renormalization} Clearly, the Poisson subalgebra
$\Symb^\reg\subset\Symb$ admits a quantization to the algebra of regular functional differential operators
with the usual formula for the product of symbols,
\begin{equation}
\begin{aligned}{}
H_1*_{Diff}H_2(\varphi,\pi)=&\exp\left(ih\int\frac{\delta}{\delta\pi_1(\x)}\frac{\delta}{\delta\varphi_2(\x)}d\x\right)\\
&H_1(\varphi_1,\pi_1)H_2(\varphi_2,\pi_2)|_{\varphi_1=\varphi_2=\varphi,\pi_1=\pi_2=\pi},
\end{aligned}
\end{equation}
or to the Moyal algebra
\begin{equation}
\begin{aligned}{}
H_1*_{Moyal}H_2(\varphi,\pi)=&\exp\frac{ih}2\int\left(\frac{\delta}{\delta\pi_1(\x)}\frac{\delta}{\delta\varphi_2(\x)}-
\frac{\delta}{\delta\pi_2(\x)}\frac{\delta}{\delta\varphi_1(\x)}\right)d\x\\
&H_1(\varphi_1,\pi_1)H_2(\varphi_2,\pi_2)|_{\varphi_1=\varphi_2=\varphi,\pi_1=\pi_2=\pi}.
\end{aligned}
\end{equation}
It is easy to see that these two quantizations are equivalent.

However, these quantizations cannot be extended to the whole $\Symb$, because the products contain pairings of
higher functional derivatives which are distributions rather than smooth functions. (Recall that in $\Symb$
only the first functional derivatives are functions.) Therefore, a product formula for $\Symb$ should be
constructed in a completely different way, it should not be a bidifferential operator. We hope that such a formula
can be found by QFT methods, cf. [6].

Assume that such a quantization $*$ has been found. Moreover, assume that we have an inclusion of quantum algebras
\begin{equation}
R:(\Symb^\reg, *_{Diff})\hookrightarrow(\Symb,*).
\end{equation}
Note that for $H\in\Symb^\reg$, we have, in general, $R(H)\ne H$. Moreover,
for a family $H_\Lambda$, $\Lambda\in\R$ of regular
Hamiltonians which tend to a non-regular Hamiltonian $H$ as $\Lambda\to\infty$, we have $R^{-1}(H_\Lambda)\to\infty$.

In Hamiltonian quantum field theory, the map $H_\Lambda\to R^{-1}(H_\Lambda)$ is called the {\it renormalization map}.

\end{document}